\documentclass[preprint,aps,amsmath,nofootinbib,tightenlines]{revtex4}
\usepackage{graphicx}
\usepackage{bm}
\usepackage{epsfig}

\newif\ifpdf
\ifx\pdfoutput\undefined
\pdffalse 
\else
\pdfoutput=1 
\pdftrue
\fi

\def\Dsl{\hbox{/\kern-.6000em D}} 

\def\dsl{\,\raise.15ex\hbox{/}\mkern-13.5mu D}
\def\bsigma{\mbox{\boldmath $\sigma$}}

\def\bsigma{\mbox{\boldmath $\sigma$}}

\def\ltap{\ \raise.3ex\hbox{$<$\kern-.75em\lower1ex\hbox{$\sim$}}\ }
\def\gtap{\ \raise.3ex\hbox{$>$\kern-.75em\lower1ex\hbox{$\sim$}}\ }
\def\OMIT#1{}

\def\lsim{\mathrel{\raise.3ex\hbox{$<$\kern-.75em\lower1ex\hbox{$\sim$}}}}
\def\gsim{\mathrel{\raise.3ex\hbox{$>$\kern-.75em\lower1ex\hbox{$\sim$}}}}

\def\msb{{\overline{\rm MS}}}

\newcommand{\nn}{\nonumber}

\newcommand{\bmk}{\mathbf k}
\newcommand{\bmp}{\mathbf p}
\newcommand{\bmq}{\mathbf q}

\newcommand{\bmA}{\mathbf A}

\newcommand{\bmsigma}{\mathbf \bsigma}

\catcode`\@=11
\def\slash{\mathpalette\make@slash}
\def\make@slash#1#2{\setbox\z@\hbox{$#1#2$}%
  \hbox to 0pt{\hss$#1/$\hss\kern-\wd0}\box0}
\catcode`\@=12 

\begin{document}
\ifpdf
\DeclareGraphicsExtensions{.pdf, .jpg}
\else
\DeclareGraphicsExtensions{.eps, .jpg}
\fi


\preprint{ \vbox{ \hbox{MPP-2003-38} 
\hbox{hep-ph/0307376}  
}}

\title{\phantom{x}\vspace{0.5cm} 
Three-Loop Anomalous Dimension of the Heavy Quark
Pair Production Current in Non-Relativistic QCD
\vspace{1.0cm} }

\author{Andr\'e H.~Hoang\vspace{0.5cm}}
\affiliation{Max-Planck-Institut f\"ur Physik\\
(Werner-Heisenberg-Institut), \\
F\"ohringer Ring 6,\\
80805 M\"unchen, Germany\vspace{1cm}
\footnote{Electronic address: ahoang@mppmu.mpg.de}\vspace{1cm}}


\begin{abstract}
\vspace{0.5cm}
\setlength\baselineskip{18pt}

The three-loop non-mixing contributions to the anomalous dimension of the
leading order quark pair production current in non-relativistic QCD are computed.  
It is demonstrated that the renormalization procedure can only be
carried out consistently if the dynamics of both soft and the ultrasoft
degrees of freedom is present for all scales below the heavy quark mass, and
if the soft and ultrasoft renormalization scales are always correlated.

\end{abstract}
\maketitle


\newpage

%
%
%
\section{Introduction}
\label{sectionintroduction}

The lineshape scan of the threshold top pair production cross
section constitutes an integral part of the top quark physics program at a
future $e^+e^-$ or $\gamma\gamma$  collider.\,\cite{TESLA,NLC,JLC} 
Because in the Standard Model the top quark width $\Gamma_t\approx 1.5$~GeV is
larger than the typical hadronization energy $\Lambda_{\rm QCD}$, it is
expected that the lineshape is a smooth function of the c.\,m.\,energy, and
that non-perturbative effects are strongly suppressed. From the rise of the 
cross section a precise measurement of the top quark mass will be
possible, while from the shape and the normalization of the cross
section one can extract the top quark Yukawa coupling $y_t$, the top
width and the strong coupling.\,\cite{TTbarsim} 
Past fixed order next-to-next-to-leading order (NNLO) computations of
the cross section have shown that a measurement of the top quark mass
in a threshold mass scheme with theoretical uncertainties of
$100$-$200$~MeV or better are feasible.\,\cite{synopsis} However, in
the fixed order approach the theoretical uncertainty of the
normalization of the NNLO cross section were estimated at the 20\%
level,\,\cite{synopsis} which would jeopardize competitive
measurements of $y_t$, $\Gamma_t$ or $\alpha_s$.
 
Recently, the renormalization-group-improved $e^+e^-$ top threshold cross
section was computed~\cite{hmst,hmst1} in the framework of an effective theory for
non-relativistic heavy quark pairs, called
vNRQCD.\,\cite{LMR,amis,amis2,amis3,hms1,hs1} 
This effective theory describes the dynamics of heavy quarkonium systems,
when the hierarchy of scales $m\gg mv\gg mv^2\gg\Lambda_{\rm QCD}$ is
satisfied, $m$ being the mass and $v$ the average c.m.\,velocity of the
quarks. The matching is carried out at the hard scale
$\mu=m$ onto a potential-like theory with both soft and ultrasoft degrees of
freedom. For scales $\mu<m$ the correlation of energy and momenta is accounted
for since the ultrasoft and soft renormalization scales $\mu_U$ and $\mu_S$
are related, $\mu_U=\mu_S^2/m\equiv m\nu^2$. The running in vNRQCD is
expressed in the dimensionless scaling parameter $\nu$. All operators (and
their coefficients) are evolved from $\nu=1$ to
$\nu\simeq v$ of order of the average c.m.\,velocity of the quarks, where
matrix elements are free from large logarithmic terms. In dimensional
regularization the factors of $\mu_U^\epsilon$ and $\mu_S^\epsilon$
multiplying operators in the renormalized effective Lagrangian are determined
uniquely from the $v$ power counting in $d$ dimensions.~\cite{amis2,hs1}

In renormalization-group-improved perturbation theory the expansion of 
the normalized cross section $R$ takes the parametric form
\begin{eqnarray}
 R \, = \, \frac{\sigma_{t\bar t}}{\sigma_{\mu^+\mu^-}}
 \, = \,
 v\,\sum\limits_k \left(\frac{\alpha_s}{v}\right)^k
 \sum\limits_i \left(\alpha_s\ln v \right)^i \times
 \bigg\{1\,\mbox{(LL)}; \alpha_s, v\,\mbox{(NLL)}; 
 \alpha_s^2, \alpha_s v, v^2\,\mbox{(NNLL)}\bigg\}
 \,,
 \label{RNNLLorders}
\end{eqnarray}
where $v\ll 1$ is the top quark velocity and 
where the indicated terms are of leading logarithmic (LL), next-to-leading
logarithmic (NLL), and next-to-next-to-leading logarithmic (NNLL)
order. In Refs.\,\cite{hmst,hmst1,hs1} all logarithms were summed in the
Wilson coefficients of the operators that contribute to the cross section at
NNLL order except for the Wilson coefficient $c_1$ of the leading order
spin-triplet current  
\begin{eqnarray} \label{J1J0}
  {\bf J}_{1,\bf p} = 
    \psi_{\bmp}^\dagger\, \bsigma (i\sigma_2) \chi_{-\bmp}^*
\,,
\end{eqnarray}
for which only the NLL anomalous dimension was known.\,\cite{LMR}. 
In Refs.\,\cite{hmst,hmst1,hs1} it was shown that the summation of logarithms
leads to a significant reduction of the normalization uncertainties, and a
theoretical  uncertainty of 3\% was estimated for the NNLL cross section in 
$e^+e^-$ annihilation. The computation of the NNLL anomalous
dimension of the current ${\bf J}_{1,\bf p}$, is an important
task for the determination of the full NNLL order cross section and
for a cross check of the error estimate made in Refs.\,\cite{hmst,hmst1,hs1}.

The evolution of $c_1(\nu)$ is obtained by integrating the anomalous dimension 
\begin{eqnarray}\label{c1anomdim}
 \nu \frac{\partial}{\partial\nu} \ln[c_1(\nu)] & = &
\gamma_{c_1}^{\rm NLL}(\nu) + \gamma_{c_1}^{\rm NNLL}(\nu) + \ldots
\,.
\end{eqnarray}
The LL order anomalous dimension is zero. The NLL order term reads~\cite{LMR} 
\begin{eqnarray} \label{c1ad}
\gamma_{c_1}^{\rm NLL}(\nu)  & = &
 -\:\frac{{\cal V}_c^{(s)}(\nu)
  }{ 16\pi^2} \bigg[ \frac{ {\cal V}_c^{(s)}(\nu) }{4 }
  +{\cal V}_2^{(s)}(\nu)+{\cal V}_r^{(s)}(\nu)
   + {\bf S}^2\: {\cal V}_s^{(s)}(\nu)  \bigg] 
   \nn\\
  && +\: \alpha_s^2(m\nu)\,\bigg[ \frac{C_F}{2}(C_F-2\,C_A)\bigg] 
     +  \alpha_s^2(m\nu)\,\bigg[  
     3 {\cal V}_{k1}^{(s)}(\nu) + 2 {\cal V}_{k2}^{(s)}(\nu) \bigg] \,,
\end{eqnarray}
where ${\bf S}^2=2$ is the squared quark total spin operator for the
spin-triplet configuration.\footnote{
In this work we keep the explicit dependence on the total quark
spin operator. Thus the results can be generalized to the
spin-singlet configuration where ${\bf S}^2=0$.
}
The terms
${\cal V}_c^{(s)}$ and ${\cal V}_2^{(s)}$, ${\cal V}_r^{(s)}$, ${\cal
  V}_s^{(s)}$ are the color singlet Wilson coefficients of the potentials of
order $\alpha_s v^{-1}$ ($1/\bmk^2$) and $\alpha_s v$ ($1/m^2$,
$(\bmp^2+\bmp^{\prime 2})/(2m^2\bmk^2)$, ${\bf S}^2/m^2$),
respectively~\cite{amis,hs1}, while ${\cal V}_{k1,k2}^{(s)}$ are 
the Color singlet Wilson coefficients of the sum operators ${\cal
  O}_{k1}^{(1)}$, ${\cal O}_{k2}^{(T)}$ 
introduced in Ref.\,\cite{hs1}. In the convention for the vNRQCD
operator basis used in Refs.\,\cite{LMR,hs1} the potentials at order $\alpha_s^2
v^0$ contained operators having the 
momentum dependence $1/|\bmk|$, $\bmk=\bmp-\bmp^\prime$ being the momentum
transfer, which mixed into $\gamma_{c_1}$ at NLL order. For technical
reasons explained below in Sec.\,\ref{subsectionsoft} we adopt in this work
the convention where all $\alpha_s^2 v^0$ potentials are represented by sum
operators in analogy to 
${\cal O}_{k1}^{(1)}$, ${\cal O}_{k2}^{(T)}$. Instead of the $1/|\bmk|$
potentials used in 
Refs.\,\cite{LMR,hs1} we thus have the sum operators ${\cal O}_k^{(1)}$ and 
${\cal O}_k^{(T)}$ giving the contribution 
$\Delta {\cal L}_p = {\cal V}_{k}^{(1)} {\cal
O}_{k}^{(1)} + {\cal V}_{k}^{(T)} {\cal O}_{k}^{(T)}$ to the vNRQCD
Lagrangian with Wilson coefficients ${\cal V}_{k}^{(1)}$ and ${\cal
  V}_{k}^{(T)}$, respectively.
At LL order the coefficients are in agreement with the 
convention in Ref.\,~\cite{hs1} and contribute to the second term on the RHS of
Eq.\,(\ref{c1ad}). The explicit form for the operators is given in the
appendix. Note that in the following we frequently refer to the sums over
field indices in the sum operators as loop integrals in order to simplify the
presentation.  

The mixing displayed in Eq.\,(\ref{c1ad}) arises from 
two-loop vertex diagrams containing only potential loops and
insertions of the Coulomb potential, the subleading heavy quark kinetic energy
operator and the $1/m$-suppressed
potentials.\,\cite{LMR} At NLL order, in Eq.\,(\ref{c1ad}),
$\alpha_s$,
${\cal V}_c^{(s)}$, 
${\cal V}_2^{(s)}$, ${\cal V}_r^{(s)}$, ${\cal V}_s^{(s)}$, 
${\cal V}_{k1,k2}^{(s)}$ and
${\cal V}_{k}^{(1,T)}$ need to be known at LL order for all values
$\nu<1$.\,\cite{amis,amis3,hms1,hs1}  
 
At NNLL order there are two classes of contributions. The first is due to the
two-loop mixing shown in Eq.\,(\ref{c1ad}) and requires the NLL results for 
$\alpha_s$, ${\cal V}_c^{(s)}$,  
${\cal V}_2^{(s)}$, ${\cal V}_r^{(s)}$, ${\cal V}_s^{(s)}$, 
${\cal V}_{k1,k2}^{(s)}$ and
${\cal V}_{k}^{(1,T)}$, but it does not modify the form
of the NLL anomalous dimension. The second class requires the
computation of three-loop vertex diagrams with potential loops and
either soft or ultrasoft loops that require new $c_1$ counterterms. We call
this class ``non-mixing contributions'' since it leads to genuinely new
contributions in the anomalous dimension of $c_1$. By power
counting there are no contributions from diagrams with three potential loops
or which have both soft and ultrasoft loops.

In this paper we present the non-mixing contributions to the NNLL anomalous
dimension of $c_1$.  As mentioned above, all order $\alpha_s^2 v^0$ potentials
are presented by sum operators. Otherwise, we use the vNRQCD velocity
renormalization group with the notations and conventions of
Refs.\,\cite{amis,amis3,hs1}. We employ the $\msb$ scheme in 
$d=4-2\epsilon$ dimensions. 


The outline of the paper is as follows:
In Sec.\,\ref{sectionmethod} we describe the method we have used to carry out
the computations and illustrate it by rederiving the NLL anomalous
dimension of $c_1$. Based on renormalization group invariance we also predict
the NNLL order $1/\epsilon^2$ term of the renormalization constant of $c_1$.
In Sec.\,\ref{sectionnonmixing} present and discuss our result for the NNLL
non-mixing contributions of the anomalous dimension of $c_1$.
Section\,\ref{subsectionusoft} and \ref{subsectionsoft} are devoted to the
contributions involving the dynamics of ultrasoft and soft degrees of freedom,
respectively. The contributions to the NNLL renormalization group equation of
$c_1$ are determined and discussed in Sec.\,\ref{subsectionsolution} and the
modifications of the formulas for the NNLL heavy quark pair production cross
section at threshold in $e^+e^-$ annihilation are discussed in
Sec.\,\ref{subsectiondecay}. 
Our conclusions are given in Sec.\,\ref{sectionconclusion}.  
The paper has three appendices where we have collected formulas for the reader
interested in the details of the computations.

\section{Method of the Computation} 
\label{sectionmethod}

The standard method to determine the renormalization constant of the 
current ${\bf J}_{1,\bf p}$  consists of computing the overall UV-divergences
of quark-antiquark-to-vacuum {\it on-shell} matrix elements of spin-triplet currents
at a certain loop order including lower order counterterm diagrams needed to 
subtract the subdivergences. There are, however, technical complications in
this approach due to the existence of IR-divergent Coulomb phases for
on-shell quarks and from the fact
that the vertex diagrams in general depend on three physical scales, the 
mass $m$, the c.\,m.\,energy $E$ and the external quark momentum
$\bmp$.
Note that imposing the on-shell condition $\bmp^2=m E$ on the quark-antiquark-to-vacuum
matrix elements in dimensional regularization, yields IR-divergent $1/\epsilon^n$
poles that are very difficult to separate from the UV-divergences. In practice
it is therefore necessary to compute the quark-antiquark-to-vacuum
matrix elements in an asymptotic expansion for $\bmp^2-m E\ll m E$, where the
IR-divergent Coulomb phases manifest themselves as powers of 
$\ln((\bmp^2-m E)/m E)$. An efficient way to avoid these complications and,
in addition, to reduce the number of 
diagrams that have to be computed is to consider current
correlator graphs rather than the vertex diagrams. The correlator graphs are
obtained from closing the external quark lines of the vertex diagrams with an
additional insertion of the current ${\bf J}_{1,\bf p}$. This means that one
has to determine diagrams with one more loop, but it reduces the
number of diagrams to be considered, eliminates the quark momentum $\bmp$ as an
external scale and avoids the IR-divergent Coulomb phases. Moreover, since all
IR-divergences cancel, using correlator graphs it is not necessary to
distinguish on- and off-shell contributions since any 
off-shell term that could be relevant for the renormalization constant leads to
scaleless integrals in dimensional regularization which automatically
vanish. In this approach the three-loop (NNLL) 
renormalization constant of the current ${\bf J}_{1,\bf p}$ is obtained from
the subdivergences in four-loop correlator diagrams
that remain after the one- and  two-loop subdivergences have been  
subtracted. The surviving  four-loop overall divergences 
are canceled by external vacuum-type diagrams and not related to the
renormalization of the current ${\bf J}_{1,\bf p}$. The method applies
analously at any order. Interestingly, the
overall divergences of the four-loop correlator graph vanish
in dimensional regularization because the graphs are non-analytic in the
c.\,m.\,energy, and there are no operators in the effective theory that could
absorb the overall divergences. The result for the renormalization constant
obtained by this method will agree with the one obtained from
quark-antiquark-to-vacuum on-shell matrix elements.

\begin{figure}[t] 
\begin{center}
 \leavevmode
 \epsfxsize=14cm
 \leavevmode
 \epsffile[45 460 545 575]{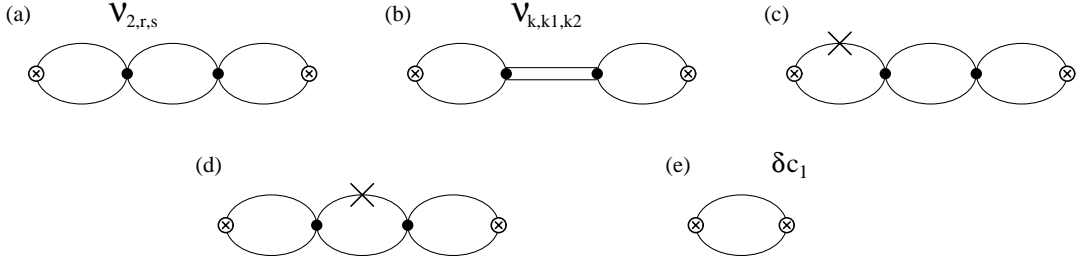}
 \vskip  0.0cm
 \caption{
Three-loop current correlator diagrams and the counterterm diagram for
the computation of the NLL anomalous dimension of $c_1$.
 \label{fignllcorrelators} }
\end{center}
\end{figure}

For illustration let us reconsider the computation of the NLL renormalization
constant that leads to Eq.\,(\ref{c1ad}). The relevant correlator diagrams are
shown in Fig.\,\ref{fignllcorrelators}.
%
%
Four-quark interactions without label refer to the Coulomb
potential. Crosses on quark lines refer to insertions of the quark kinetic
energy at subleading order. Here and throughout the paper, combinatorial
factors and diagrams obtained by flipping the graphs left-to-right and
up-to-down are to be understood. There are no one-loop subdivergences that
have to be subtracted in this case. Adopting the form
\begin{equation}
c_1^0 \, = \, c_1 \, + \, \delta c_1
 \, = \, Z_{c_1}\,c_1
\end{equation}
for the unrenormalized Wilson coefficient of the current ${\bf J}_{1,\bf p}$ 
with the renormalization constant written as
\begin{equation}
Z_{c_1} \, = \, 1 \, + \,
\frac{\delta z_{c_1}^{\rm NLL}}{\epsilon}\, + \,
\Big(\,
\frac{\delta z_{c_1}^{\rm NNLL,2}}{\epsilon^2} \, + \,
\frac{\delta z_{c_1}^{\rm NNLL,1}}{\epsilon}\,\Big)\, + \,
\ldots
\,,
\end{equation}
one obtains
\begin{eqnarray}
\delta z_{c_1}^{\rm NLL} & = & \frac{1}{4}\,\gamma_{c_1}^{\rm NLL}(\nu)
\,,
\end{eqnarray}
where $\gamma_{c_1}^{\rm NLL}(\nu)$ is given in Eq.\,(\ref{c1ad}).
Taking into account the velocity renormalization group equations for the
vNRQCD coefficients in $d=4-2\epsilon$ dimensions 
($\nu \frac{d}{d\nu} g_s(m\nu)=-\epsilon g_s(m\nu)+\ldots$, 
$\nu \frac{d}{d\nu} g_s(m\nu^2)=-2\epsilon g_s(m\nu^2)+\ldots$, etc.)
one can derive Eq.\,(\ref{c1ad}) using that $c_0$ is
renormalization group invariant. From the same relation and the fact that all
$1/\epsilon^n$ ($n=1,2,\ldots$) terms cancel in the renormalization group
equations one also finds the coefficient of the $1/\epsilon^2$ term at NNLL
order ($a_s\equiv\alpha_s(m\nu)$, $a_u\equiv\alpha_s(m\nu^2)$),
\begin{eqnarray}\label{delta22predict}
\delta z_{c_1}^{\rm NNLL,2} & = &
-\frac{a_s^2 a_u}{24\pi}\,C_F\,(2 C_F^2 + 3 C_A C_F + C_A^2) 
-\frac{a_s^2}{192 \pi^2}\, C_F \,\beta_0\,( V_2^{(s)} + V_r^{(s)} + {\bf S}^2
                     V_s^{(s)})
\nonumber\\ & &  
+ \frac{a_s^3}{48 \pi}\,\beta_0\,\bigg\{\,  
      C_A C_F 
    + \frac{C_F^2}{4}\bigg[\,3 - c_D + 2 c_F^2\Big(1 - \frac{2}{3}{\bf S}^2\Big)\,\bigg]
    - 2(3{\cal V}_{k1}^{(s)} +2 {\cal V}_{k2}^{(s)})
\nonumber\\ & & \hspace{2cm}  
    - C_F (\,C_{2a}^{(2)} - C_F(C_{2b}^{(2)} + 2C_{2c}^{(2)})\,)
   \,\bigg\}
\nonumber\\ & &  
+ \frac{a_s^3}{288\pi}\,C_F\,  \bigg\{ 
      8 C_A^2 + 14 C_F^2 
      + C_A C_F\bigg[\, \frac{29}{2} + \frac{11}{2} c_D 
           + c_F^2(-13 + 7{\bf S}^2) \,\bigg] \,\bigg\}
\,,
\end{eqnarray}
where
$\beta_0=11/3 C_A-4/3T n_l$ is the one-loop QCD beta-function,
and $c_D$ and $c_F$ are the coefficients of the Darwin and the magnetic
operators of the HQET action~\cite{Bauer1}. 
The terms $C_{2a,2b,2c}^{(2)}$ are coefficients associated to the soft 6-field   
operators ${\cal O}_{2\varphi,2A,2c}^{(2)}$, which were introduced in
Ref.\,\cite{hs1}, and which mix into the order $\alpha_s v$ potential
operators. The formulas for the coefficients, which all depend on the scaling
parameter $\nu$, are collected in App.\,\ref{appendixC}. 
Note that at NNLL order only the diagrams that lead to non-mixing
contributions of the anomalous dimension of $c_1$ can also contribute to
$\delta Z_{c_1}^{\rm NNLL,2}$. Thus, Eq.\,(\ref{delta22predict}) provides a
non-trivial cross check for our result and the consistency of the effective
theory under renormalization.   

\section{Non-mixing contributions at NNLL order} 
\label{sectionnonmixing}

For the non-mixing contributions  to the NNLL anomalous
dimension of $c_1$ one can distinguish between four-loop correlator diagrams
containing ultrasoft loops and those with soft loops. Diagrams with both
ultrasoft and soft loops do not exist at this order.  
Note that the ultrasoft and soft non-mixing contributions are separately
gauge-invariant since they are induced by ultrasoft and soft gluons,
respectively, which represent different degrees of freedom in the effective
theory. 

\subsection{Ultrasoft non-mixing contributions at NNLL order} 
\label{subsectionusoft}
%
%

%
%
\begin{figure}[t] 
\begin{center}
 \leavevmode
 \epsfxsize=14cm
 \leavevmode
 \epsffile[45 245 545 575]{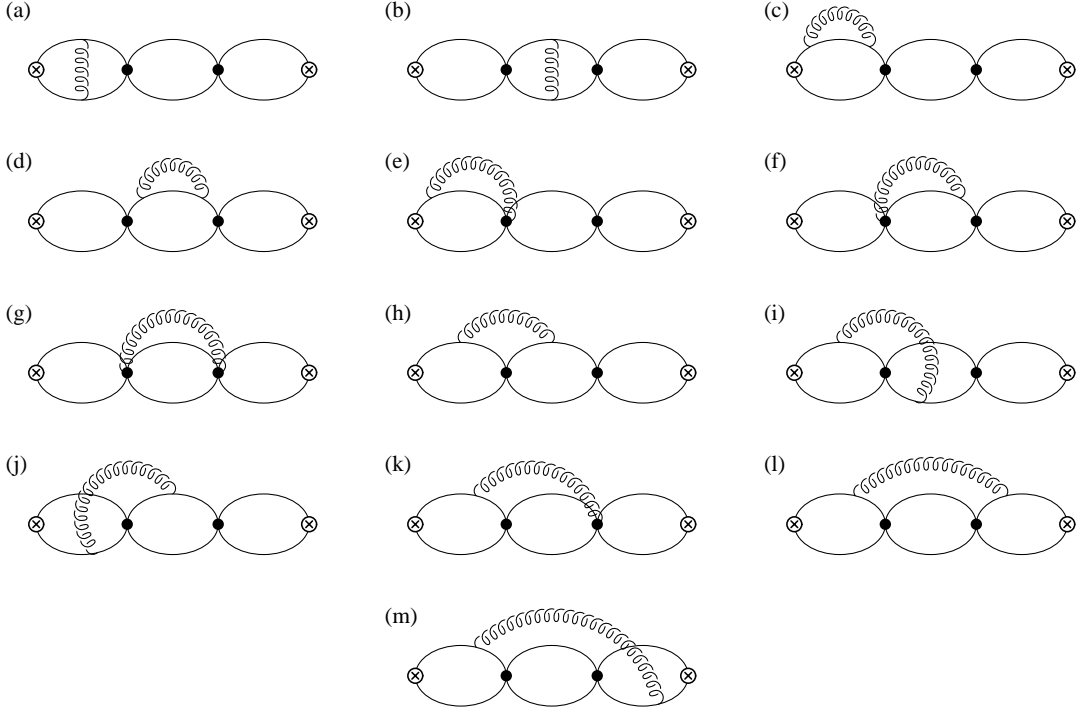}
 \vskip  0.0cm
 \caption{
Four-loop graphs for the calculation of the ultrasoft non-mixing contributions of
the NNLL anomalous dimension of $c_1$.
 \label{figusoftcorrelators} }
\end{center}
\end{figure}
%
%
%
%
\begin{figure} 
\begin{center}
 \leavevmode
 \epsfxsize=14cm
 \leavevmode
 \epsffile[45 390 545 575]{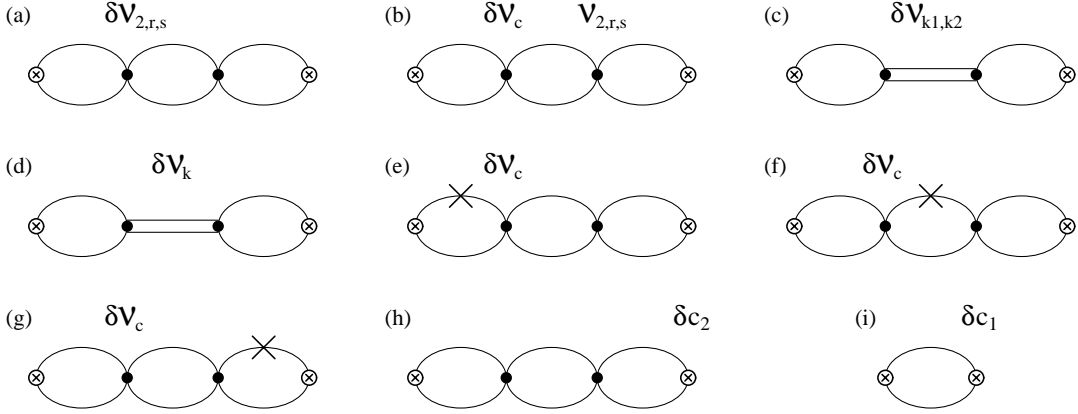}
 \vskip  0.0cm
 \caption{
Counterterm graphs for the removal of subdivergences in graphs of
Figs.\,\ref{figusoftcorrelators} and \ref{figsoftcorrelators}.
Graphs with wave function renormalization constants are to be understood.
 \label{figcounter} }
\end{center}
\end{figure}

For the determination of the ultrasoft non-mixing contributions we adopted
the Coulomb gauge, where the propagation of the longitudinal ultrasoft gluon
field component $A^0$ does not contribute. For the transverse ultrasoft gluon
field  $\bmA$  the leading order couplings to quarks and the Coulomb potential 
have to be taken into account. The complete set of four-loop correlator
diagrams is displayed in  
Figs.\,\ref{figusoftcorrelators}. The diagrams needed to cancel the one-loop
subdivergences are displayed in Figs.\,\ref{figcounter}a,c,h
using the graphical notation from Refs.\,\cite{LMR,amis,hs1}.
Here, $\delta c_2$ denotes the counterterm of the Wilson coefficient
of the $v^2$-suppressed spin-triplet current 
$1/m^2\,\psi_{\bmp}^\dagger\,\bmp^2 \bsigma
(i\sigma_2) \chi_{-\bmp}^*$.~\cite{hmst,hmst1}
There are no two-loop subdivergences.
After the removal of the one-loop subdivergences the contribution to  
the renormalization factor $Z_{c_1}$ from each of the diagrams in
Figs.\,\ref{figusoftcorrelators} is of the form
\begin{eqnarray}
\frac{\alpha_s(\mu_U)\,\alpha_s^2(\mu_S)}{4\pi}\,\bigg[\,
\frac{A}{\epsilon^2}\, + \, \frac{B}{\epsilon}
\,\bigg]
\,.
\label{c1usoftpieces}
\end{eqnarray} 
The contributions to $A$ and $B$ as well
as the totals are given in Tab.\,\ref{tabc1}. The results coming from 
graphs \ref{figusoftcorrelators}l and \ref{figusoftcorrelators}m are zero, but  
non-trivial and have been obtained from explicit computation.    
\begin{table}[t!]
\begin{center}
\begin{tabular}{c|c|c}
\hline\hline
 \mbox{\hspace{2.2cm}}Graph\mbox{\hspace{2.2cm}} & 
 \mbox{\hspace{2.2cm}}$A$\mbox{\hspace{2.2cm}} & 
 \mbox{\hspace{2.2cm}}$B$\mbox{\hspace{2.2cm}}\\
 \hline
 a & $0$                       & $-\frac{2}{3}C_F^3$ \\[2mm]
 b & $-\frac{1}{3}C_F^3$       & $-\frac{2}{3}(\ln 2+\frac{1}{6})C_F^3$ \\[2mm]
 c & $0$                       & $-\frac{2}{3}C_F^3$ \\[2mm]
 d & $-\frac{1}{3}C_F^3$       & $-\frac{2}{3}(\ln 2+\frac{1}{6})C_F^3$ \\[2mm]
 e & $0$                       & $-C_A C_F^2$ \\[2mm]
 f & $-\frac{1}{2}C_A C_F^2$   & $-(\ln 2+\frac{1}{2})C_A C_F^2$ \\[2mm]
 g & $-\frac{1}{12}C_A^2 C_F$  & $-\frac{1}{6}(\ln 2+\frac{7}{6})C_A^2 C_F$ \\[2mm]
 h & $-\frac{1}{6}C_F C_1$     & $-\frac{5}{3}(\ln 2-\frac{1}{6})C_F C_1$ \\[2mm]
 i & $-\frac{1}{12}C_F C_1$    & $-\frac{5}{6}(\ln 2-\frac{1}{6})C_F C_1$ \\[2mm]
 j & $-\frac{1}{12}C_F C_1$    & $-\frac{5}{6}(\ln 2-\frac{1}{6})C_F C_1$ \\[2mm]
 k & $-\frac{1}{6}C_A C_1$     & $-\frac{5}{3}(\ln 2-\frac{1}{6})C_A C_1$ \\[2mm]
 l & $0$                       & $0$ \\[2mm]
 m & $0$                       & $0$ \\[2mm]
 \hline  
 Total &
  $-\frac{1}{6}C_F( 2 C_F^2+3 C_A C_F + C_A^2)$ &
  $C_F ( 2 C_F^2 - C_A C_F - C_A^2 )\,\ln 2$ \\[1mm]
 & &
  $-\frac{1}{18} C_F ( 38 C_F^2 + 27 C_A C_F + C_A^2 )$\\[2mm]
 \hline\hline
\end{tabular}
\end{center}
{\caption{
Contributions to Eq.\,(\ref{c1usoftpieces}) from the graphs in
Figs.\,\ref{figusoftcorrelators}. 
}
\label{tabc1} }
\end{table}
The results do not involve factors
of $\rho\equiv\gamma_E-\ln(4\pi)$ because we have implemented the $\msb$ scheme
by scaling each term $\mu_S$ or $\mu_U$ with a factor $e^{\rho/2}$.
Note that each counterterm diagram can in general contribute to
several four-loop diagrams. For example, the counterterm diagrams involving  
$\delta {\cal V}_{2,r,s}$ are required for the subtraction of one-loop
subdivergences in the four-loop graphs
\ref{figusoftcorrelators}a--\ref{figusoftcorrelators}d and 
\ref{figusoftcorrelators}h--\ref{figusoftcorrelators}j, whereas the diagrams 
with $\delta {\cal V}_{k1,k2}$ have to be considered for the graphs 
\ref{figusoftcorrelators}f, \ref{figusoftcorrelators}g and
\ref{figusoftcorrelators}k. On the other hand, the diagram with 
$\delta c_2$ is needed for 
the graphs \ref{figusoftcorrelators}a and \ref{figusoftcorrelators}c. The
graphs \ref{figusoftcorrelators}e, \ref{figusoftcorrelators}l and
\ref{figusoftcorrelators}m do not have any one- or two-loop  
subdivergences. From the totals given in Tab.\,\ref{tabc1} one finds agreement
with the prediction of the ultrasoft $1/\epsilon^2$ term made in
Eq.\,(\ref{delta22predict}). 

It is an interesting conceptual aspect that the divergences in the graphs in
Figs.\,\ref{figusoftcorrelators} can only be renormalized 
consistently if the correlation of soft and ultrasoft scales,
$\mu_U=\mu_S^2/m=m\nu^2$ is taken into account (see also
Ref.\,\cite{hmst1}). As an example, let us consider the 
contribution to $\delta Z_{c_1}^{\rm NNLL,2}$ induced by the diagram 
Fig.\,\ref{figusoftcorrelators}g in some detail. 
The result for the diagram reads
($E\equiv-(\sqrt{s}-2m+i\epsilon)$, $p^2\equiv m E$)
\begin{eqnarray} \label{graphg}
 {\rm Fig.\,\ref{figusoftcorrelators}g} 
& = & 
-i\,C_A^2\,C_F \,\frac{\alpha_s(\mu_U)\alpha_s^2(\mu_S)}{4\pi} 
\,\frac{m\,p}{4\pi}\,
\nonumber\\[2mm]
& &\mbox{\hspace{1cm}}\times
\bigg[\, \frac{1}{6\epsilon^2} + 
 \frac{1}{\epsilon}\,\bigg(
   \frac{1}{6}\,\ln\Big(\frac{\mu_U^2}{E^2}\Big)
  +\frac{1}{2}\,\ln\Big(\frac{\mu_S^2}{p^2}\Big)
  -2\ln 2 + \frac{29}{18}\,\bigg)
\,+\, \ldots\,
\bigg] 
\,.
\end{eqnarray}
The corresponding counterterm contribution from the graph in
Fig.\,\ref{figcounter}c reads
\begin{eqnarray} \label{graphcounter}
 {\rm Fig.\,\ref{figcounter}c} 
& = & 
-i\,C_A^2\,C_F \,\frac{\alpha_s(\mu_U)\alpha_s^2(\mu_S)}{4\pi} 
\,\frac{m\,p}{4\pi}\,
\nonumber\\[2mm]
& &\mbox{\hspace{1cm}}\times
\bigg[\, -\frac{1}{3\epsilon^2} + 
 \frac{1}{\epsilon}\,\bigg(
  -\ln\Big(\frac{\mu_S^2}{p^2}\Big)
  +2\ln 2 - \frac{7}{3}\,\bigg)
\,+\, \ldots\,
\bigg] 
\,
\end{eqnarray}
and the sum including the $c_1$ counterterm graph from Fig.\,\ref{figcounter}i
gives 
\begin{eqnarray} \label{graphsum}
\lefteqn{
{\rm Fig.\,\ref{figusoftcorrelators}g}
+{\rm Fig.\,\ref{figcounter}c} 
+{\rm Fig.\,\ref{figcounter}i} 
\, = \,
-i\,C_A^2\,C_F \,\frac{\alpha_s(\mu_U)\alpha_s^2(\mu_S)}{4\pi} 
\,\frac{m\,p}{4\pi}\,
\bigg[\, -\frac{1+12 A\delta}{6\epsilon^2}
}
\nonumber\\[2mm]
& & 
 + \, \frac{1}{\epsilon}\,\bigg(
   \frac{1}{6}\,\ln\Big(\frac{\mu_U^2}{E^2}\Big)
  -\Big(\frac{1}{2}+2 A\delta\Big)\,\ln\Big(\frac{\mu_S^2}{p^2}\Big)
  +4\,A\delta\,\Big(\ln 2-1\Big) 
  -2\,B\delta
  -\frac{13}{18} 
\,\bigg)
\,+\, \ldots\,
\bigg] 
\,,
\end{eqnarray}
where $\delta=(C_A^2 C_F)^{-1}$.
This leads to the following contribution for the $c_1$ counterterm
\begin{eqnarray} \label{ABresults}
A & = & -\,\frac{1}{12}\,C_A^2\,C_F
\,,
\nonumber\\[2mm]
B & = &
\frac{1}{6}\,\bigg[\,
\ln\Big(\frac{m \, \mu_U}{\mu_S^2}\Big)
-\ln 2-\frac{7}{6}
\,\bigg]\,C_A^2\,C_F
\,.
\end{eqnarray}
The dependence on $\ln\mu_S$ and $\ln\mu_U$ in $B$ vanishes only if  
the correlation $\mu_U=\mu_S^2/m$ is accounted for in
Fig.\,\ref{figusoftcorrelators}g as well as in the counterterm graphs needed
to cancel the subdivergences.  
This is the case for all graphs in Figs.\,\ref{figusoftcorrelators} that
contribute to $B$. This
demonstrates that the correlation of the soft and the ultrasoft renormalization
scales, which is given unambiguously from the mass dimension and the $v$
counting of the operators (see Ref.\,\cite{amis2,hs1}), is also needed to
ensure the renormalizability and consistency of the effective theory. It
also shows that
the correlation of the soft and ultrasoft scales for all scales below $m$ is
an integral property of the effective theory for nonrelativistic dynamic
(i.e. non-static) heavy quark pairs.

\subsection{Soft non-mixing contributions at NNLL order} 
\label{subsectionsoft}

\begin{figure}[t] 
\begin{center}
 \leavevmode
 \epsfxsize=14cm
 \leavevmode
 \epsffile[45 390 545 575]{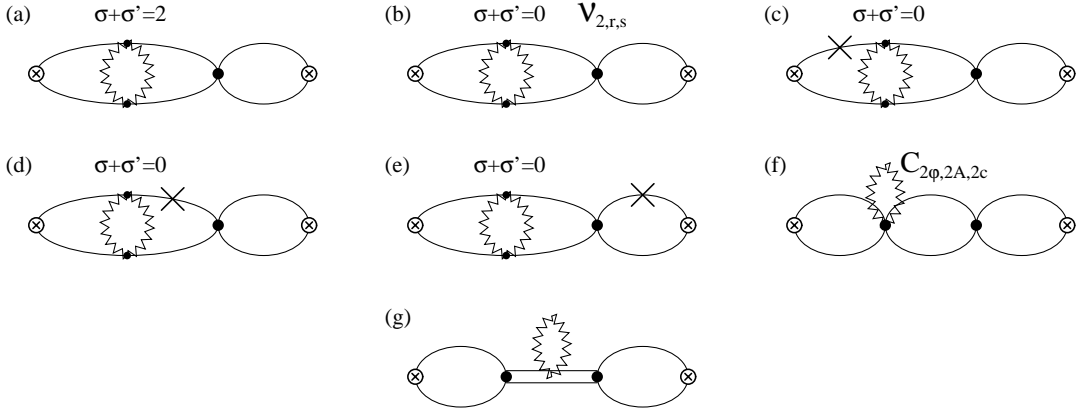}
 \vskip  0.0cm
 \caption{
Four-loop graphs for the calculation of the soft non-mixing contributions of
the NNLL anomalous dimension of $c_1$.
 \label{figsoftcorrelators} }
\end{center}
\end{figure}

\begin{figure}[t] 
\begin{center}
 \leavevmode
 \epsfxsize=14cm
 \leavevmode
 \epsffile[45 525 545 575]{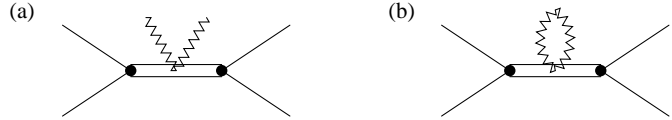}
 \vskip  0.0cm
 \caption{
(a) Six-field operators that contribute to the soft running of  
${\cal O}_{k,k1,k2}^{1,T}$ through the graphs in (b).  
 \label{fignewsumop} }
\end{center}
\end{figure}

The four-loop correlator diagrams relevant for the soft non-mixing
contributions are displayed in Figs.\,\ref{figsoftcorrelators}, where we have
used again the graphical notations of Refs.\,\cite{LMR,amis,hs1}. 
%
%
Diagrams with insertions of two soft vertices involving the functions
$U^{(\sigma)}$, $W^{(\sigma)}$, $Y^{(\sigma)}$,
$Z^{(\sigma)}$~\cite{amis,amis2}  
are shown in Figs.\,\ref{figsoftcorrelators}a-e. Insertions of the 6-field
operators ${\cal O}_{2\varphi,2A,2c}^{(2)}$ from Ref.\,\cite{hs1} are shown in 
Fig.\,\ref{figsoftcorrelators}f. In Fig.\,\ref{figsoftcorrelators}g there are
also insertions of soft 6-field sum operators (see Fig.\,\ref{fignewsumop}a),
which, upon closing the soft lines  (Fig.\,\ref{fignewsumop}b), contribute to
the soft running of the operators ${\cal O}_{k,k1,k2}^{1,T}$. (The ultrasoft
running of the operators ${\cal O}_{k,k1,k2}^{1,T}$ was discussed in detail in 
Ref.\,\cite{hs1}.)
%
%
The four-quark matrix elements of these
operators, with the intermediate sum and the soft loop integration being
carried out, are given in App.\,\ref{appendixB}. 
For a part of these matrix elements, as explained in App.\,\ref{appendixB},
we used results obtained earlier in 
Ref.\,\cite{Kniehl1} based on the asymptotic expansion of QCD diagrams.
To ensure consistency of the effective theory under renormalization this
enforces the convention for the operators ${\cal O}_{k}^{(1,T)}$ explained in
Sec.\,\ref{sectionintroduction}. 
The diagrams needed to cancel the one-loop soft subdivergences are displayed
in Figs.\,\ref{figcounter}a--g. There are no two-loop subdivergences. 
After the removal of the one-loop subdivergences the contribution to
the renormalization factor $Z_{c_1}$ from each of the diagrams in
Figs.\,\ref{figsoftcorrelators} can be written in the form
\begin{eqnarray}
\frac{\alpha_s^3(\mu_S)}{4\pi}\,\bigg[\,
\frac{A^\prime}{\epsilon^2}\, + \, \frac{B^\prime}{\epsilon}
\,\bigg]
\,.
\label{c1softpieces}
\end{eqnarray} 
The contributions to $A^\prime$ and $B^\prime$ are given in Tab.\,\ref{tabc2}.  
We note that the total spin operator for the quarks is defined as 
${\bf S}= \frac{1}{2}({\bmsigma}_1 + {\bmsigma}_2)$ using three-dimensional
Pauli matrices. This is the common convention of
non-relativistic quantum mechanics and has also been adopted for the vNRQCD
action~\cite{amis,amis2}. We also note that other schemes for the quark spin
in the effective theory are possible.
\begin{table}[t!]
\begin{center}
\begin{tabular}{c|c|c}
\hline\hline
 Graphs & $A^\prime$ & $B^\prime$  \\
 \hline
 a & $\frac{7}{36}C_F^3+\frac{5}{24}C_A C_F^2(1-\frac{1}{45}c_F^2(6+{\bf S}^2))$ 
   & $\frac{5}{6}C_F^3-\frac{1}{18}C_A C_F^2(11-\frac{17}{36}c_F^2(3-4{\bf S}^2))$ \\[2mm]
   & $-\frac{1}{12}C_F^2 T n_l(1-\frac{1}{3}c_D+\frac{2}{9}c_F^2(3-2{\bf S}^2))$
   & $+\frac{1}{6}C_F^2 T n_l(1+\frac{1}{9}c_D-\frac{2}{27}c_F^2(3-2{\bf S}^2))$ \\[4mm]
 b & $-\frac{1}{48 a_s \pi}\beta_0 C_F(
             {\cal V}_2^{(s)}+{\cal V}_r^{(s)}+{\bf S}^2{\cal V}_s^{(s)})$       
   & $\frac{1}{36 a_s \pi}(\beta_0-4C_A)C_F(
             {\cal V}_2^{(s)}+{\cal V}_r^{(s)}+{\bf S}^2{\cal V}_s^{(s)})$ \\[4mm]
 c+d+e 
   & $\frac{1}{24}\beta_0 C_F^2$                       
   & $-\frac{1}{18}(\beta_0-4 C_A)C_F^2$ \\[4mm]
 f & $-\frac{1}{12}\beta_0 C_F(C_{2a}^{(2)}- C_F C_{2b}^{(2)}-2 C_F C_{2c}^{(2)})$       
   & $-\frac{1}{18}(\beta_0+8C_A)C_F(C_{2a}^{(2)}- C_F C_{2b}^{(2)})$ \\[2mm]
   & 
   & $-\frac{2}{9}(\beta_0-4 C_A)C_F^2C_{2c}^{(2)}$ \\[4mm]
 g & $\frac{1}{12}\beta_0C_F(C_A-\frac{1}{2}C_F)+\frac{1}{9}C_AC_F(2C_F+C_A)$
   & $\frac{1}{18}\beta_0C_F(C_F-\frac{37}{8}C_A)
      +\frac{1}{9}C_AC_F(C_F+\frac{101}{16}C_A)$ \\[2mm]
   & $-\frac{1}{6}\beta_0(3{\cal V}_{k1}^{(s)}+2{\cal V}_{k2}^{(s)})$
   & $+\frac{1}{6}(7\beta_0-16C_A){\cal V}_{k1}^{(s)}
      +\frac{1}{36}(25\beta_0-64C_A){\cal V}_{k2}^{(s)}$ \\[2mm]
 \hline\hline
\end{tabular}
\end{center}
{\caption{
Contributions to Eq.\,(\ref{c1softpieces}) from the graphs in
Fig.\,\ref{figsoftcorrelators}. 
}
\label{tabc2} }
\end{table}

Like for the ultrasoft contributions, we find agreement with the prediction of
the soft $1/\epsilon^2$ terms made in Eq.\,(\ref{delta22predict}). This
demonstrates the consistency of the vNRQCD action under renormalization
at non-trivial subleading order. In particular, the agreement shows that the
dynamics of soft and ultrasoft degrees of freedom are both needed
simultaneously for {\it all\,} scales below heavy quark mass $m$.

\subsection{Solution of the anomalous dimension at NNLL order} 
\label{subsectionsolution}

From the results shown in Tabs.\,\ref{tabc1} and \ref{tabc2} we find the
following result for the non-mixing contributions of the NNLL anomalous
dimension of $c_1$ ($a_s\equiv\alpha_s(m\nu)$, $a_u\equiv\alpha_s(m\nu^2)$),
\begin{eqnarray}
\gamma_{c_1,nm}^{\rm NNLL}(\nu) & = &
\frac{a_s^2 a_u}{\pi}\,\Big[\,
2\,C_F ( 2 C_F^2 - C_A C_F - C_A^2 )\,\ln 2
\,-\,\frac{1}{9}\,C_F ( 38 C_F^2 + 27 C_A C_F + C_A^2 )
\,\Big]
\nn \\ & &
+\:\frac{a_s^2}{24 \pi^2}\,(\beta_0 - 4C_A)\,C_F\, 
  (\,V_2^{(s)} + V_r^{(s)} + {\bf S}^2 V_s^{(s)}\,)
\nn \\ & & 
- \: \frac{a_s^3}{12\pi}\,(\beta_0 + 8 C_A)\,C_F\,
     (\,C_{2a}^{(2)} - C_F ( C_{2b}^{(2)} - 4 C_{2c}^{(2)} )) 
+  4 \frac{a_s^3}{\pi}\, C_A \,C_F^2\,C_{2c}^{(2)} 
\nn \\ & & 
+ \: \frac{a_s^3}{4\pi}\,(7 \beta_0 - 16 C_A){\cal V}_{k1}^{(s)} 
 + \frac{a_s^3}{24\pi}\,(25\beta_0 - 64 C_A)\,{\cal V}_{k2}^{(s)}
\nn \\ & &
 -\: \frac{a_s^3}{48\pi}\,\beta_0\,\bigg\{\,   
 \frac{37}{2} C_A \,C_F + C_F^2\,\bigg[\,9 + c_D 
         - \frac{2}{3} c_F^2\Big(3 - 2\,{\bf S}^2\Big)\,\bigg]\,\bigg\} 
\nn \\ & &
+\: \frac{a_s^3}{48\pi}\,\bigg\{\, \frac{101}{2}\, C_A^2\, C_F 
   + C_A C_F^2\,\bigg[\, 13 + \frac{11}{3} c_D 
       - \frac{1}{3}c_F^2\,\Big(5 + 8{\bf S}^2\Big)\,\bigg] 
       + 60 C_F^3\,\bigg\}
\,.
\label{gammaNNLLnonmix}
\end{eqnarray}
Let us write the solution of Eq.\,(\ref{c1anomdim}) for $\nu< 1$ as
\begin{eqnarray}
 \ln\Big[ \frac{c_1(\nu)}{c_1(1)} \Big] & = &
\xi^{\rm NLL}(\nu) + 
\Big(\,
\xi^{\rm NNLL}_{\rm m}(\nu) + \xi^{\rm NNLL}_{\rm nm}(\nu)
\,\Big) + \ldots
\,.
\label{c1solution}
\end{eqnarray}
The NLL order term $\xi^{\rm NLL}(\nu)$ was determined in
Refs.\,\cite{hs1,Pineda2}. From Eq.\,(\ref{gammaNNLLnonmix}) we find the
following form for the NNLL non-mixing term,
\begin{eqnarray}
\xi^{\rm NNLL}_{\rm nm}(\nu) & = &
b_2\,\alpha_s(m)^2\,(1 - z) 
+ b_3\,\alpha_s(m)^2\,(1 - z^2) 
+ b_4\,\alpha_s(m)^2\,\bigg[\,1 - z + 2\ln(w)\,\bigg] 
\nn \\[2mm] & &
+\: b_5\,\alpha_s(m)^2\,
      \bigg[\,\frac{5}{2} - 2z - \frac{1}{2} z^2 + (4 - z^2)\ln(w)\,\bigg]  
\nn \\[2mm] & &
+\: b_6\,\alpha_s(m)^2\,\bigg[\,1 - z^{2 - 2 CA/\beta_0}\,\bigg] 
+ b_7\,\alpha_s(m)^2\,\bigg[\,1 - z^{2 - 13 CA/(6\beta_0)}\,\bigg] 
\,, 
\label{xiNNLLnm}
\end{eqnarray}
where the coefficients $b_i$ read
\begin{eqnarray} 
b_2 & = & \frac{ C_F^2 (\beta_0 - 4C_A)}
              {6 \beta_0^2 (6\beta_0 - 13 C_A)(\beta_0 - 2C_A)}\,
 \Big\{ 6 \beta_0^2 \Big[ 2C_F + C_A({\bf S^2} - 3)\Big] 
\nn \\ & & \hspace{2cm}
        + \beta_0 C_A \Big[-74 C_F + C_A (42 - 13 {\bf S^2})\Big] 
        +  C_A^2 (100 C_F - 9 C_A)\Big\}\,,
\nn \\[2mm] 
b_3 & = & \frac{C_F}{3744 C_A \beta_0^2}\,
 \Big\{ - 3 \beta_0^2 \Big[ 481 C_A^2 + 350 C_A C_F - 64 C_F^2\Big] 
\nn \\ & & \hspace{2cm}
        + \beta_0 C_A \Big[ 3939 C_A^2 + 2474 C_A C_F + 2648 C_F^2\Big] 
        + 48 C_A^2 C_F \Big[100 C_F - 9 C_A\Big]\Big\}\,,
\nn \\[2mm] 
b_4 & = & \frac{2 C_F}{9 \beta_0}\,
 \Big\{ C_A^2 \Big[ 1 + 18 \ln 2 \Big] 
        + 9 C_A C_F \Big[ 3 + 2 \ln 2 \Big] 
        - 2 C_F^2 \Big[ -19 + 18 \ln 2 \Big] \Big\}\,,
\nn \\[2mm] 
b_5 & = &  \frac{C_F}{18 \beta_0^2}\,
 \Big\{ \beta_0 \Big[ 17 C_A^2 + 30 C_A C_F + 4 C_F^2 \Big] 
        - 32 C_A \Big[ C_A^2 + 3 C_A C_F + 2 C_F^2 \Big] \Big\}\,,
\nn \\[2mm]
b_6 & = & \frac{C_F^2}{ 144 (\beta_0 - 2 C_A)(\beta_0 - C_A)}\,
 \Big\{ \beta_0^2\Big[-3 + 4{\bf S^2}\Big] 
        + 6 \beta_0 C_A \Big[ 9 - 10 {\bf S^2} \Big] 
\nn \\ & & \hspace{2cm} 
        + 4 C_A^2 \Big[ -33 + 32 {\bf S^2} \Big] \Big\}\,,
\nn \\[2mm]
b_7 & = & -\: \frac{C_F^2 (5C_A + 8C_F)}
                  {39 C_A (6\beta_0 - 13 C_A) (12 \beta_0 - 13 C_A)}\,
 \Big\{ 18 \beta_0^2 - 171 \beta_0 C_A + 385 C_A^2) \Big\}
\,.
\label{bcoeffs}
\end{eqnarray}
Note that $b_4$ originates exclusively from the ultrasoft corrections
determined in Sec.\,\ref{subsectionusoft}.

One can also determine the NNLL mixing contributions 
of the anomalous dimension at the hard scale $\nu=1$ since the corresponding
matching conditions of the couplings appearing in Eq.\,(\ref{c1ad}) are
available.\,\cite{amis2,hs1} The result reads
\begin{eqnarray}
\gamma_{c_1,m}^{\rm NNLL}(\nu=1) & = &
\frac{\alpha_s(m)^3}{48 \pi}\, C_F^2\, 
\bigg[\, C_A \,\Big(16\,{\bf S}^2-3\Big) 
     + 4\, C_F\,\Big(5 - 2\,{\bf S}^2\Big) - \frac{16}{5}\, T \,\bigg]
\,.
\label{gammaNNLLmix}
\end{eqnarray}
From Eq.\,(\ref{gammaNNLLmix}) we can determine the $\alpha_s(m)^3\ln\nu$ term of
the NNLL mixing contribution, which is the first term in the expansion in
terms of $\alpha_s(m)$,
\begin{eqnarray}
\xi^{\rm NNLL}_{\rm m}(\nu) & = &
\gamma_{c_1,m}^{\rm NNLL}(1)\,\ln\nu 
+ {\cal O}(\alpha_s^4\ln^2\nu)
\,.
\label{xiNNLLm}
\end{eqnarray}
Numerically the NNLL non-mixing contributions to $c_1$ are rather large and
entirely dominated by the first term in Eq.\,(\ref{gammaNNLLnonmix}) which
originates from the ultrasoft corrections.
In Tab.\,\ref{tabcompare} the values for $\xi^{\rm NLL}(\nu)$ and
$\xi^{\rm NNLL}_{\rm nm}(\nu)$ are displayed for different $\nu$ for the top
and the bottom quarks. For top quarks we find that the NNLL non-mixing
contributions are of the same size as the NLL terms for the relevant region
$\nu\sim v\approx 
0.2$. Here, the new NNLL order corrections shift $c_1$ by about
$+5\%$. However, it is not yet possible to draw phenomenological conclusions
for the normalization of the top threshold cross section in $e^+e^-$
collisions from this result, since the NNLL mixing contributions for $\nu<1$
are still unknown. Because these corrections involve two ultrasoft loop
integrations arising in the NLL running of the 
coefficients in Eq.\,(\ref{c1ad}), they could potentially be large as
well. It is therefore an important task to determine the mixing corrections
for all scales below $m$.

For bottom quarks the NNLL non-mixing contributions are more than five
times larger than the NLL terms for the relevant region $\nu\sim v\approx
0.3$--$0.4$. This is not unexpected because for bottomonium systems the
binding energy $\sim m v^2$ is already of order $\Lambda_{\rm QCD}$. Thus our
result seems to affirm that for $b\bar b$ states non-perturbative effects have
a rather strong influence, and that the vNRQCD description ceases to work even
for the ground state.\,\cite{hs1} However, also for the bottom quark case the 
knowledge of the NNLL mixing contributions might be useful to gain further
insight into this issue.  
\begin{table}[tbh]
\begin{center}
\begin{tabular}{|c||c|c||c|c|}
\hline
 & \multicolumn{2}{|c||}{$m=175$\,GeV}
 & \multicolumn{2}{|c|}{$m=4.8$\,GeV} \\ \hline 
 $\quad\nu\quad$ 
 &  $\quad\xi^{\rm NLL}(\nu)\quad$ 
 & $\quad\xi^{\rm NNLL}_{\rm nm}(\nu)\quad$
 &  $\quad\xi^{\rm NLL}(\nu)\quad$ 
 & $\quad\xi^{\rm NNLL}_{\rm nm}(\nu)\quad$ \\ \hline\hline
 $1.0$ & $0.0000$ & $0.0000$ & $0.0000$ & $0.0000$ \\ \hline
 $0.9$ & $0.0033$ & $0.0019$ & $0.0145$ & $0.0181$ \\ \hline
 $0.8$ & $0.0069$ & $0.0041$ & $0.0308$ & $0.0425$ \\ \hline
 $0.7$ & $0.0110$ & $0.0069$ & $0.0495$ & $0.0770$ \\ \hline
 $0.6$ & $0.0157$ & $0.0104$ & $0.0712$ & $0.1304$ \\ \hline
 $0.5$ & $0.0211$ & $0.0150$ & $0.0968$ & $0.2254$ \\ \hline
 $0.4$ & $0.0274$ & $0.0216$ & $0.1335$ & $0.4537$ \\ \hline
 $0.3$ & $0.0349$ & $0.0318$ &  &  \\ \hline
 $0.2$ & $0.0435$ & $0.0512$ &  &  \\ \hline
\end{tabular}
\end{center}
{\caption{
Numerical values for  $\xi^{\rm NLL}(\nu)$ and
$\xi^{\rm NNLL}_{\rm nm}(\nu)$. The values for $m$ are pole masses. 
The numbers are obtained by evaluation of the analytic results using 
four-loop running for $\alpha_s$ and taking 
$\alpha_s^{(5)}(175\,\mbox{GeV})=0.107$ and 
$\alpha_s^{(4)}(4.8\,\mbox{GeV})=0.216$ 
as input.
}
\label{tabcompare} }
\end{table}
\subsection{Production Cross Section at Threshold and Comparison}
\label{subsectiondecay}

Due to the convention we use for the order $\alpha_s v^0$ potentials, our 
formulas for the NNLL order vector current correlator
${\cal A}_1$, which is used to express the quark pair production cross section
in $e^+e^-$ annihilation, slightly differs from Ref.\,~\cite{hs1}. This also
affects the two-loop 
matching condition for $c_1$. The difference arises because the corresponding
matrix elements are UV-divergent and the d-dependent  
contributions that arise from summing the intermediate indices in the
operators  ${\cal O}_k^{1,T}$ in dimensional regularization (see
App.\,\ref{appendixA}) lead to  
modifications of the UV-finite terms. Altogether, we now need four different
types of corrections to the current correlator, $\delta G^{k}_{\rm CACF}$,
$\delta G^{k}_{\rm CF2}$,  $\delta G^{k1}$ and $\delta G^{k2}$ 
to account for the corrections originating from the order $\alpha_s^2 v^0$
potentials. The results for $\delta G^{k1}$ and $\delta G^{k2}$ were given in 
Ref.\,\cite{hs1}. The corrections $\delta G^{k}_{\rm CACF}$ and
$\delta G^{k}_{\rm CF2}$ arise from the terms proportional to $C_AC_F$ and
$C_F^2$, respectively, in the color singlet combination of the operators
${\cal O}_k^{1}$ and ${\cal O}_k^{T}$ and have the form
\begin{eqnarray}
 \delta G^{k}_{\rm CACF}(a,v,m,\nu) 
 & = & 
 -\frac{m^2}{8 \pi\, a}\left\{\,
 i\,v - a\left[\,\ln\left(\frac{-i\,v}{\nu}\right)
 -\frac{5}{4}+\ln 2+\gamma_E+\Psi\left(1-\frac{i\,a}{2\,v}\right)\,\right]
 \,\right\}^2
\label{deltaGkn1}
\nn \\ & &
 +\,\frac{m^2}{8 \pi\,a}\,\left[\, -v^2 + 
 \frac{a^2}{16}\,\left(\frac{1}{\epsilon^2}
 -\frac{3}{\epsilon} - 11\right)
 \,\right]
\,,
 \nonumber \\[3mm]
 \delta G^{k}_{\rm CF2}(a,v,m,\nu) 
 & = & 
 -\frac{m^2}{8 \pi\, a}\left\{\,
 i\,v - a\left[\,\ln\left(\frac{-i\,v}{\nu}\right)
 -1+\ln 2+\gamma_E+\Psi\left(1-\frac{i\,a}{2\,v}\right)\,\right]
 \,\right\}^2
\label{deltaGkn2}
\nn \\ & &
 +\,\frac{m^2}{8 \pi\,a}\,\left[\, -v^2 + 
 \frac{a^2}{16}\,\left(\frac{1}{\epsilon^2}
 -\frac{2}{\epsilon} - 12\right)
 \,\right]
\,.
\end{eqnarray}
The vector current correlator ${\cal A}_1$ at NNLL order then reads
\begin{eqnarray}
{\cal A}_1(v,m,\nu)
 & = &
 6 \,N_c\,\Big[\,
 G^c(a^\prime,v,m,\nu)
 + \Big({\cal{V}}_2^{(s)}(\nu)+2{\cal{V}}_s^{(s)}(\nu)\Big)\, 
 \delta G^\delta(a,v,m,\nu) 
\nonumber \\[2mm] & &
+ \:{\cal{V}}_r^{(s)}(\nu)\,\delta G^r(a,v,m,\nu) 
+ \delta G^{\rm kin}(a,v,m,\nu) 
\nn \\[2mm] & &
- \:C_AC_F\,\alpha_s^2(m\nu)\,\delta G^k_{\rm CACF}(a,v,m,\nu) 
+\frac{C_F^2}{2}\,\alpha_s^2(m\nu)\,\delta G^k_{\rm CF2}(a,v,m,\nu)
\nn \\[2mm] & &
+ \:\alpha_s^2(m\nu){\cal V}_{k1}^{(s)}(\nu)\,\delta G^{k1}(a,v,m,\nu) 
+   \alpha_s^2(m\nu){\cal V}_{k2}^{(s)}(\nu)\,\delta G^{k2}(a,v,m,\nu)
\,\Big]
\,,\quad\nn
\\[2mm]
a & = & -\frac{1}{4\,\pi}\,{\cal V}_{c}^{(s)}(\nu)\,,
\qquad
a^\prime \, = \, -\frac{1}{4\,\pi}\,{\cal V}_{c,{\rm eff}}^{(s)}(\nu)
\,,
\end{eqnarray}
where $G^c$, $\delta G^\delta$, $\delta G^r$, $\delta G^{\rm kin}$ and
$\delta G^{k1}$, $\delta G^{k2}$, ${\cal V}_{c}^{(s)}$, ${\cal V}_{c,{\rm
    eff}}^{(s)}$ were 
given in Refs.\,\cite{hmst1} and \cite{hs1}, respectively. The two-loop
matching condition for $c_1$ now 
reads 
\begin{eqnarray} \label{c1match}
c_1(1) & = & 1- \frac{2 C_F}{\pi}\: {\alpha_s(m)} +   
  \alpha_s^2(m) \bigg[C_F^2\bigg(\frac{\ln 2}{3}-\frac{31}{24}
  -\frac{2}{\pi^2}\bigg) + C_A C_F\bigg(\frac{\ln 2}{2}-\frac{5}{8}\bigg) 
  + \frac{\kappa}{2} \bigg]
\,,\quad
\end{eqnarray}
where the constant $\kappa$ was determined in Refs.\,\cite{Hoang1}.

Recently, the order $\alpha_s^3\ln\alpha_s$ corrections to the heavy
quarkonium partial width into a lepton pair were computed using an asymptotic 
expansion of QCD diagrams close to threshold\,\,\cite{Kniehl2}. In this work
the summation of higher order logarithms was not attempted.
The results were partly based on three-loop
quark-antiquark-to-vacuum matrix elements in the off-shell limit $m E\neq 0$,
$p^2=0$\,\cite{Peninprivate}. With the result for the total cross section of 
quark pair production at threshold in $e^+e^-$ annihilation in
Refs.\,\cite{hmst,hmst1,hs1}, and including the new formulas given above,
the order $\alpha_s^3\ln\alpha_s$ corrections can also be
derived by expanding out the summations contained in the Wilson coefficients
for energies on the bound state poles. 
The contributions proportional to ${\bf S}^2=S(S+1)$ in our result 
do not agree with Ref.~\cite{Kniehl2} because there the terms $S(S+1)$ 
do not refer to the three-dimensional quark spin
operators~\cite{Peninprivate} and are defined in a different scheme for the
quark spin. 
For the $\alpha_s^3\ln\alpha_s$ corrections to the heavy
quarkonium partial width into a lepton pair we find a 
discrepancy, which would correspond to the term 
$\frac{1}{3}C_A^2C_F-\frac{4}{3}C_AC_F^2(2-5\ln 2)$
being added to the constant ${\cal C}_1$ of Ref.\,\cite{Kniehl2}.
This term has the same sign and approximately the same size as the constant 
${\cal C}_1$ itself and affects the error estimates made in
Ref.\,\cite{Kniehl2}. 
%

%
%
\section{Conclusion}
\label{sectionconclusion}

In this work we have determined, within the framework of vNRQCD, the
non-mixing contributions to the NNLL anomalous dimension of the leading
spin-triplet current ${\bf J}_{1,\bf p}$, which describes the production of
non-relativistic quark-antiquark pair in $e^+e^-$ annihilation. 
Our result for the $1/\epsilon^2$ terms of the
three-loop renormalization constant of the current is consistent with the
constraints of renormalization group invariance imposed on the already known
NLL UV-divergences. This demonstrates that both soft and ultrasoft
degrees of freedom need to be present in the effective theory for all scales 
below the heavy quark mass $m$ as predicted by vNRQCD. Moreover, a sequence of 
different effective theories with an intermediate uncorrelated matching scale
is not consistent under renormalization. Our results also show that the NNLL
order renormalization 
constant of the current is independent of the renormalization scales for the
soft and ultrasoft degrees of freedom, only if their correlation is accounted
for, as given unambiguously from the mass dimension and the $v$
counting of the operators in vNRQCD. 
We have derived an updated expression for the NNLL total cross section of
heavy quark pair production at threshold in $e^+e^-$ annihilation.

\begin{acknowledgments} 
We would like to thank A.\,Manohar and I.\,Stewart for many useful discussions
and for comments to the manuscript. We thank A.~Penin for a comparison of some
intermediate results that were obtained in
Ref.\,\cite{Kniehl2}. We particularly thank the Aspen Center for Physics and
also the Benasque Center of Science, where part of this work was carried out.
\end{acknowledgments}

\appendix

\section{Convention for the potential operators ${\cal O}_k^{(1,T)}$}
\label{appendixA}

The order $\alpha_s^2 v^0$ potential sum operators ${\cal O}_k^{(1,T)}$ used
in this work are obtained by matching to the one-loop $Q\bar Q$ scattering
amplitude in full QCD in the threshold expansion and carrying out only the
$dq^0$ integration of the loop momentum $q=(q^0,\bmq)$ for the contributions
of order $\alpha_s^2 v^0$. The explicit expression for the operators is 
($g_s=g_s(m \nu)$)
\begin{eqnarray} \label{Ok1T}
{\cal O}_{k}^{(1)} & = & 
  -\: \frac{g_s^4 \tilde\mu_S^{4\epsilon}}{4m}\: 
 \sum_{{\bf p,p',q}}\,C_1 \Big[\, g_0 - g_2 \,\Big]
 \big[ \psi_{\bmp^\prime}^\dagger \psi_{\bmp}
 \chi_{-\bmp^\prime}^\dagger \chi_{-\bmp} \big] 
\,, \nn\\
{\cal O}_{k}^{(T)} & = & 
  - \:\frac{g_s^4 \tilde\mu_S^{4\epsilon}}{4m}\: 
 \sum_{{\bf p,p',q}}\,\Big[
  -\frac{1}{4}(C_d-C_A)\,g_0 + C_A\, g_1 + \frac{1}{4}(C_d+C_A)\,g_2
 \,\Big] 
 \big[ \psi_{\bmp^\prime}^\dagger T^A \psi_{\bmp}
 \chi_{-\bmp^\prime}^\dagger \bar T^A \chi_{-\bmp} \big] 
\,, \nn\\
\end{eqnarray}
where the functions $g_i$ have the form
\begin{eqnarray}
 g_0 & = & \Big[(\bmq-\bmp)^2+(\bmq-\bmp^\prime)^2-(\bmp-\bmp^\prime)^2\Big]\,
          \frac{(\bmq-\bmp)^2+(\bmq-\bmp^\prime)^2}
               {(\bmq-\bmp)^4\,(\bmq-\bmp^\prime)^4}
    \,, \nn\\[2mm]
 g_1 & = & \frac{(\bmq-\bmp)^2+(\bmq-\bmp^\prime)^2+(\bmp-\bmp^\prime)^2}
            {(\bmq-\bmp)^2\,(\bmq-\bmp^\prime)^2\,(\bmp-\bmp^\prime)^2} 
    \,, \nn\\[2mm]
 g_2 & = & \frac{\bmq^2-\bmp^2}{(\bmq-\bmp)^4\,(\bmq-\bmp^\prime)^2}
     + \frac{\bmq^2-\bmp^{\prime 2}}{(\bmq-\bmp)^2\,(\bmq-\bmp^\prime)^4}
  \,.
\end{eqnarray}
They give the contribution $\Delta {\cal L}_p = {\cal V}_{k}^{(1)} {\cal 
O}_{k}^{(1)} + {\cal V}_{k}^{(T)} {\cal O}_{k}^{(T)}$ to the vNRQCD
Lagrangian. The Wilson coefficients read 
\begin{equation}
{\cal V}_{k}^{(1)}(\nu) \, = \,
{\cal V}_{k}^{(T)}(\nu) \, = \, 1
\,.
\label{Vk1Tone}
\end{equation}
Note that the tilde over the renormalization scales $\mu_U$ and $\mu_S$ refers
to the $\msb$ definition  $\tilde\mu_{U,S}=e^{\rho/2}\,\mu_{U,S}$, where
$\rho\equiv\gamma_E-\ln(4\pi)$.
When ${\cal O}_k^{(1,T)}$ are inserted in four-quark matrix
elements or used in the Schr\"odinger equation with the sum over the
intermediate index $\bmq$ being
carried out in dimensional regularization, they take at leading order the form 
\begin{eqnarray} \label{Ok1Texplicit}
\langle i {\cal O}_{k}^{(1)}\rangle & = & 
  - \:i\,\frac{g_s^4 \tilde\mu_S^{4\epsilon}}{4m (\bmk^2)^{(2-n/2)}}\,
 C_1\, \Big[\, f(1,1)-f(2,1) \,\Big]\,I\otimes\bar I
\,, \nn\\
\langle i {\cal O}_{k}^{(T)}\rangle & = & 
  - \:i\,\frac{g_s^4 \tilde\mu_S^{4\epsilon}}{4m (\bmk^2)^{(2-n/2)}}\,
 \,\Big[\, \Big(\frac{5}{2}C_A-2C_F\Big)\,f(1,1) -
           \Big(\frac{3}{2}C_A-2C_F\Big)\,f(2,1)
 \,\Big] 
 \,T^A\otimes\bar T^A \,, \nn\\
\end{eqnarray}
where
\begin{eqnarray}
f(a,b) 
& = &
\frac{
\Gamma(a+b-\frac{n}{2})\Gamma(\frac{n}{2}-a)\Gamma(\frac{n}{2}-b)
}{\Gamma(a)\Gamma(b)\Gamma(n-a-b)(4\pi)^{\frac{n}{2}}} 
\,,
\end{eqnarray}
and $\bmk=\bmp-\bmp^\prime$. Here $n=d-1=3-2\epsilon$.
The matrix element of the color singlet combination ${\cal O}_{k}^{(s)}={\cal
  O}_{k}^{(1)}-C_F{\cal O}_{k}^{(T)}$ reads
\begin{eqnarray}
\langle i {\cal O}_{k}^{(s)}\rangle & = & 
-\:8i\,\frac{\pi^2\alpha_s^2(m\nu)\tilde\mu_S^{4\epsilon}}{m (\bmk^2)^{(2-n/2)}}\,
\bigg[\,\frac{1-n}{2}\,C_A\,C_F + \frac{n-2}{2}\,C_F^2
\,\bigg]\,f(1,1)
\,,
\label{Oksinglet}
\end{eqnarray}
which agrees with Ref.\,\cite{Beneke1}.

\section{4-quark matrix elements of 6-field operators}
\label{appendixB}

The soft running of the operators ${\cal O}_k^{(1,T)}$, ${\cal O}_{k_1}^{(1)}$ and
${\cal O}_{k_2}^{(T)}$ originates from four-quark matrix elements of 
6-field sum operators (see Fig.\,\ref{fignewsumop}a) in close analogy to the operators 
${\cal O}_{2\varphi,2A,2c}^{(2)}$, which contribute to the soft running of the
spin-independent order $\alpha_s v$ ($1/m^2$, $(\bmp^2+\bmp^{\prime
  2})/(2m^2\bmk^2)$, ${\bf S}^2/m^2$) potentials.\,\cite{hs1} 
The divergences in these 4-quark matrix elements have already been deduced
earlier, but for the computation of the NNLL anomalous dimension of $c_1$
their full form is required. 
To be definite we call the 6-field operators
$\tilde{\cal O}_k^{(1,T)}$, $\tilde{\cal O}_{k_1}^{(1)}$ and
$\tilde{\cal O}_{k_2}^{(T)}$.
The operators $\tilde{\cal O}_k^{(1,T)}$ and $\tilde{\cal O}_{k_1}^{(1)}$,
$\tilde{\cal O}_{k_2}^{(T)}$ are responsible for the soft running of the operators 
${\cal O}_k^{(1,T)}$ and ${\cal O}_{k_1}^{(1)}$, ${\cal O}_{k_2}^{(T)}$,
respectively. They give the contribution 
$\Delta {\cal L}_p = \tilde{\cal V}_{k}^{(1)} \tilde{\cal 
O}_{k}^{(1)} + \tilde{\cal V}_{k}^{(T)} \tilde{\cal O}_{k}^{(T)}+
\tilde{\cal V}_{k1}^{(1)}\tilde{\cal O}_{k1}^{(1)}+
\tilde{\cal V}_{k2}^{(T)}\tilde{\cal O}_{k2}^{(T)}$ to the vNRQCD
Lagrangian.  The operators $\tilde{\cal O}_k^{(1)}$ and
$\tilde{\cal O}_k^{(T)}$ arise from a matching computation at two loops
similar to the one 
described in App.\,\ref{appendixA} for ${\cal O}_k^{(1,T)}$. In analogy to
Eq.\,(\ref{Vk1Tone}) one finds that  
$\tilde{\cal V}_{k}^{(1)}(\nu)=\tilde{\cal V}_{k}^{(T)}(\nu)=1$. The leading
order 4-quark matrix element for the unrenormalized color-singlet combination
of $\tilde {\cal O}_k^{(1,T)}$ with the sum over intermediate indices and the 
soft loop being carried out (Fig.\,\ref{fignewsumop}b)
can be derived in an expansion in $\epsilon$
from results given Ref.\,\cite{Kniehl1} and reads
\begin{eqnarray}
\langle i(\tilde{\cal O}_k^{(1)}-C_F \tilde{\cal O}_k^{(T)})\rangle & = &
 i \,\frac{\pi\alpha_s(m\nu)^3 \tilde\mu_S^{2\epsilon}\mu_S^{4 \epsilon}}
               {m (\bmk^2)^{7/2 - n}}\,C_F\,
 \bigg\{\, 
      \frac{1}{\epsilon}\,\bigg[\,\frac{\beta_0}{4}\,(2C_A - C_F) 
                        +  \frac{2}{3}C_A (C_A + 2C_F) \,\bigg]
\nn \\ & & \hspace{1cm}
      + \beta_0\,\bigg[\, C_A\bigg(\frac{25}{48} + \ln 2\bigg) 
                        + C_F\bigg(\frac{1}{3} - \frac{1}{2}\ln 2 \bigg) \,\bigg] 
\nn \\ & & \hspace{1cm}
      - C_A^2\bigg( \frac{15}{16} - \frac{4}{3} \ln 2 \bigg) 
      - C_A C_F \bigg(3 - \frac{8}{3}\ln 2\bigg)
+ {\cal O}(\epsilon)
 \,\bigg\}
\,,
\end{eqnarray}
which is sufficient for the determination of the renormalization constant for
$c_1$. The divergent $\beta_0$ term is responsible for the running of
$\alpha_s$ contained in the definition of ${\cal O}_k^{(s)}$ 
(Eq.\,(\ref{Oksinglet})), while the other divergence is related to the
evolution of ${\cal V}_{k1}^{(1)}$ and ${\cal V}_{k2}^{(T)}$.

The renormalization of  $\tilde{\cal O}_{k_1}^{(1)}$ and 
$\tilde{\cal O}_{k_2}^{(T)}$ is in complete analogy to the one of the
operators ${\cal O}_{k_1}^{(1)}$ and ${\cal O}_{k_2}^{(T)}$. Thus one finds 
$\tilde{\cal V}_{k1}^{(1)}(\nu)={\cal V}_{k1}^{(1)}(\nu)$ and 
$\tilde{\cal V}_{k2}^{(T)}(\nu)={\cal V}_{k2}^{(T)}(\nu)$ at LL order, and the
coefficients vanish at the hard scale. The leading order four quark 
matrix elements have the form
\begin{eqnarray}
\langle i \tilde{\cal O}_{k1}^{(1)}\rangle & = &
  -\:2^{6 - n}\,i \,\frac{\alpha_s^3(m\nu)\pi^4\tilde\mu_S^{6\epsilon}}
                     {m (\bmk^2)^{7/2 - n}}\, 
        \frac{(1 - n)^2\,(12 - 6n + n^2)}{n(n - 2)\,\cos(\frac{n \pi}{2})}\,
  \Big[\, C_A\,(4n - 1) - 4 \,T n_l\, \Big]
\nn \\ & & \hspace{1cm}
\times f\Big(\frac{5 - n}{2}, 1\Big)\,f\Big(\frac{n - 2}{2}, 1/2\Big)
\,I\otimes\bar I 
\,,
\nn \\[2mm]
\langle i \tilde{\cal O}_{k2}^{(T)}\rangle & = &
 -\:2^{4 - n}\,i\,\frac{\alpha_s^3(m\nu)\pi^4\tilde\mu_S^{6\epsilon}}
                       {m (\bmk^2)^{7/2 - n}}\,
       \frac{(1 - n)^2\,(17 - 9n + 2n^2)}{ n(n - 2)\,\cos(\frac{n \pi}{2})}\,
    \Big[\, C_A\,(4n - 1) - 4 \,T n_l \,\Big]
\nn \\ & & \hspace{1cm}
\times f\Big(\frac{5 - n}{2}, 1\Big)\,f\Big(\frac{n - 2}{2}, 1/2\Big)
\,T^A\otimes\bar T^A 
\,.
\end{eqnarray}
The divergences are responsible for the running of the coefficient
${\cal V}_c^{(T)}$ contained in the definition of ${\cal O}_{k1}^{(1)}$ and 
${\cal O}_{k2}^{(T)}$.\,\cite{hs1}

\section{Collection of Wilson coefficients}
\label{appendixC}

The gauge invariant HQET coefficients that appear in this work are~\cite{Bauer1}
\begin{eqnarray}
  c_F(\nu) = z^{-C_A/\beta_0}\,, \quad
  c_D(\nu) = z^{-2C_A/\beta_0} +\bigg(\frac{20}{13}+\frac{32 C_F}{13 C_A}\bigg)
  \big[ 1 - z^{-13C_A/(6\beta_0)} \big]  
\,.
\end{eqnarray}
For the color singlet channel the coefficients of the order $\alpha_s v$
potentials relevant for our result are~\cite{amis,hs1}
\begin{eqnarray} \label{Vsrslts}
 {\cal V}_r^{(s)}(\nu) &=& -4\pi\,C_F\,\alpha_s(m)\,z\, \Big[ 1 - 
  \frac{8 C_A}{3\beta_0}\,\ln(w) \Big]\,, \nn\\[5pt]
 {\cal V}_2^{(s)}(\nu) &=& \pi C_F \,\alpha_s(m)\,\left(z - 1\right) 
 \left[ \frac{33}{13} +\frac{32 C_F}{13 C_A}+ \frac{9C_A}{13\beta_0} 
 -\frac{100C_F}{13\beta_0}\right] \nn \\[3pt]
 &-& \frac{8 \pi C_F(3\beta_0\! -\! 11C_A)(5C_A \!+\! 8C_F)}
  {13\, C_A(6\beta_0\! -\! 13 C_A)}\,\alpha_s(m)\,
  \left[z^{1 - (13C_A)/(6\beta_0)} - 1\right] \nn\\[5pt] 
 &-& \frac{\pi\, C_F (\beta_0 \!-\! 5C_A)}{(\beta_0 \!-\! 2C_A)}\,\alpha_s(m)\,
 \left[z^{1 - 2C_A/\beta_0} - 1\right] 
 -\frac{16\pi C_F(C_A\!-\! 2C_F)}{3\beta_0}\,\alpha_s(m)\,z \ln(w) \,,\nn\\[5pt]  
{\cal V}_s^{(s)}(\nu) &=& \frac{-2 \pi C_F}{(2 C_A-\beta_0) } \,\alpha_s(m)\,
  \bigg[ C_A + \frac{1}{3} ( 2\beta_0 - 7 C_A) \ z^{(1-2 C_A/\beta_0)} \bigg]
  \, \,,
\end{eqnarray}
whereas for the order $\alpha_s v^0$ potentials they read
\begin{eqnarray} \label{Vk1Vc1}
  {\cal V}_{k1}^{(s)}(\nu) &=& \frac{8C_A C_1}{3\beta_0}\: \ln(w)\,,
  \qquad\qquad
  {\cal V}_{k2}^{(s)}(\nu) \, = \, \frac{2C_AC_F(C_A+C_d)}{3\beta_0}\: \ln(w)
\,. 
\end{eqnarray}
The coefficients associated to the 6-field operators ${\cal
  O}_{2\varphi,2A,2c}^{(2)}$ read~\cite{hs1}
\begin{eqnarray} \label{C2slnQCD}
 C_{2a}^{(2)}(\nu) & = & \frac{4C_1}{3\beta_0}
    \ln(w)
  \,,\qquad\quad
 C_{2b}^{(2)}(\nu) =  \frac{3C_A\!-\!C_d\!-\!4C_F}{3\beta_0}
    \ln(w)
   \,,\nn\\[2mm]
 C_{2c}^{(2)}(\nu) &=& \frac{-4C_A}{3\beta_0}\, 
  \ln(w)
 \,.
\end{eqnarray}
For the results we used the definitions
\begin{eqnarray} \label{zw}
  z=\frac{ \alpha_s(m\nu) }{ \alpha_s(m) }\,,\qquad\quad
  w=\frac{ \alpha_s(m\nu^2) }{ \alpha_s(m\nu) } \,.
\end{eqnarray}
For SU($N_c$)-QCD the color coefficients that appear are 
\begin{eqnarray}
 C_A=N_c \,,\quad C_F=\frac{N_c^2-1}{2N_c} \,,\quad 
 T=\frac{1}{2} \,,\quad
 C_d=8C_F-3C_A \,, \quad C_1=\frac{1}{2}\,C_F C_A-C_F^2\,. 
\end{eqnarray}


\end{document}